\documentclass[sigconf]{acmart}
\AtBeginDocument{%
  \providecommand\BibTeX{{%
    \normalfont B\kern-0.5em{\scshape i\kern-0.25em b}\kern-0.8em\TeX}}}

\acmConference[ACM-BCB 2020]{11th ACM Conference on Bioinformatics, Computational Biology, and Health Informatics (ACM BCB)}{August 30--September 2}{Virtual}
\usepackage{hyperref}
\begin{document}

\title[GPU Docking on OLCF Summit]{GPU-Accelerated Drug Discovery with Docking on the Summit Supercomputer: Porting, Optimization, and Application to COVID-19 Research}

\author{Scott LeGrand}
\affiliation{%
  \institution{NVIDIA Corporation}
 \city{Santa Clara}
 \state{California}
 }
%\email{slegrand@corporation.com}

\author{Aaron Scheinberg}
\affiliation{%
  \institution{Jubilee Development}
%  \streetaddress{1 Th{\o}rv{\"a}ld Circle}
 \city{Cambridge}
 \state{Massachusetts}
 % \country{Iceland}
 }
%\email{aaron@jubileedev.com}
 
\author{Andreas F. Tillack}
\orcid{0000-0002-1832-3030}
\affiliation{%
  \institution{Scripps Research}
  \city{La Jolla}
   \state{California}
}

\author{Mathialakan Thavappiragasam}
\orcid{0000-0003-1240-3475}
\affiliation{%
  \institution{Oak Ridge National Laboratory}
  \city{Oak Ridge}
   \state{Tennessee}
}

\author{Josh V. Vermaas}
\affiliation{%
  \institution{Oak Ridge National Laboratory}
  \city{Oak Ridge}
   \state{Tennessee}
}
\orcid{0000-0003-3139-6469}
%\email{vermaasjv@ornl.gov}

\author{Rupesh Agarwal}
\orcid{0000-0003-1029-2281}
\affiliation{%
  \institution{University of Tennessee, Knoxville}
 % \streetaddress{30 Shuangqing Rd}
  \city{Knoxville}
  \state{Tennessee}
 % \country{China}
  }

\author{Jeff Larkin}
\orcid{0000-0001-7132-3120}
%\email{webmaster@marysville-ohio.com}
\affiliation{%
  \institution{NVIDIA Corporation}
 \city{Santa Clara}
  \state{California}
}

\author{Duncan Poole}
\affiliation{%
  \institution{NVIDIA Corporation}
  \city{Santa Clara}
  \state{California}
 }
 
\author{Diogo Santos-Martins}
\orcid{0000-0003-4622-3747}
\affiliation{%
  \institution{Scripps Research}
%  \streetaddress{8600 Datapoint Drive}
  \city{La Jolla}
  \state{California}
%  \postcode{78229}
}

\author{Leonardo Solis-Vasquez}
\affiliation{%
  \institution{TU Darmstadt}
%  \streetaddress{8600 Datapoint Drive}
  \city{Darmstadt}
  \country{Germany}
%  \postcode{78229}
}

\author{Andreas Koch}
\affiliation{%
  \institution{TU Darmstadt}
%  \streetaddress{8600 Datapoint Drive}
  \city{Darmstadt}
  \country{Germany}
%  \postcode{78229}
}

\author{Stefano Forli}
\orcid{0000-0002-5964-7111}
\affiliation{%
  \institution{Scripps Research}
%  \streetaddress{8600 Datapoint Drive}
  \city{La Jolla}
  \state{California}
%  \postcode{78229}
}

\author{Oscar Hernandez}
\affiliation{\institution{Oak Ridge National Laboratory}
\city{Oak Ridge}
\state{Tennessee}
 }

\author{Jeremy C. Smith}
\affiliation{\institution{UT/ORNL}
\city{Oak Ridge}
\state{Tennessee}
 }
\affiliation{\institution{University of Tennessee}
\city{Knoxville}
\state{Tennessee}
 }

\author{Ada Sedova}
\orcid{0000-0002-8233-3057}
\affiliation{%
  \institution{Oak Ridge National Laboratory}
  \city{Oak Ridge}
   \state{Tennessee}
}
\email{sedovaaa@ornl.gov}

\renewcommand{\shortauthors}{LeGrand et al.}

\begin{abstract}
Protein-ligand docking is an \textit{in silico} tool used to screen potential drug compounds for their ability to bind to a given protein receptor within a drug-discovery campaign. Experimental drug screening is expensive and time consuming, and it is desirable to carry out large scale docking calculations in a high-throughput manner to narrow the experimental search space. Few of the existing computational docking tools were designed with high performance computing in mind. Therefore, optimizations to maximize use of high-performance computational resources available at leadership-class computing facilities enables these facilities to be leveraged for drug discovery. Here we present the porting, optimization, and validation of the AutoDock-GPU program for the Summit supercomputer, and its application to initial compound screening efforts to target proteins of the SARS-CoV-2 virus responsible for the current COVID-19 pandemic.\footnote{This manuscript has been authored by UT-Battelle, LLC under Contract No. DE-AC05-00OR22725 with the U.S. Department of Energy. The United States Government retains and the publisher, by accepting the article for publication, acknowledges that the United States Government retains a non-exclusive, paid- up, irrevocable, world-wide license to publish or reproduce the published form of this manuscript, or allow others to do so, for United States Government purposes. The Department of Energy will provide public access to these results of federally sponsored research in accordance with the DOE Public Access Plan (\url{http://energy.gov/ downloads/doe-public-access-plan}).
}

\end{abstract}





\maketitle

\section{Introduction}
The binding of a drug to a protein target \textit{in vivo} can elicit a molecular response, such as the inhibition of a cellular function, and forms the basis of targeted drug discovery. Computational protein-ligand docking can be utilized for the rapid structure-based screening of small molecule drug candidates for target binding \cite{Forli2016,Pagadala2017}. Three dimensional structural models of both a protein receptor and a set of small molecule compounds, or ligands, are employed to computationally predict the ability of a ligand to bind to the receptor, using an optimization algorithm within some function, which can be either a physics-based empirical energy potential or statistical. \cite{Guedes2018,Trott2009,Morris2009}. Advances in high-throughput experimental screening, both cell-based \cite{Nierode2016,shen2019high} or molecular \cite{Goddard2004,jin2020structure}, have allowed tens of thousands of chemical compounds to be tested simultaneously. However, these assays are expensive, and the ability to computationally filter chemical compounds for their propensity to bind to a target protein can significantly reduce the overall cost and time to solution. The data to be evaluated are plentiful as the relevant chemical search space consists of billions of compounds. For instance, the Enamine REAL database (\url{https://enamine.net}) contains over a billion small molecules, and the ZINC 15 database contains 230 million ready-to-dock compounds, \cite{Sterling2015}. Thus the number of compounds that can potentially be screened with \textit{in silico} docking to complement experimental efforts is now on the order of billions. Increases in processor speeds and the number of cores on large nodes of computing clusters and cloud resources have made meeting this docking challenge theoretically attainable. Unfortunately, many docking programs are not inherently designed with massive high-throughput screens in mind. Most open source docking programs commonly applied in academic research are CPU-based, single-node, and utilize file-based input and output \cite{Trott2009,Morris2009}. Such codes are often designed to perform a single ligand-docking calculation per run instance of the executable. 

A variety of programs exist for ligand docking, with some using empirical but physics-based potential energy descriptions and others using fully empirical statistical potentials trained on a set of experimental structures of ligands bound to proteins \cite{Morris2009,Trott2009,Morrone2020}. The widely employed AutoDock program saw its inception in 1990 \cite{goodsell1990automated} and has undergone several major changes, with the latest version being AutoDock4. This program incorporates a physics-based energy function that includes entropic and solvation terms, in addition to van der Waals terms, hydrogen bond, and electrostatic interactions \cite{Morris2009,Morris1998}. The energy minimization procedure consists of a Lamarckian genetic algorithm (LGA), whereby a local search accompanies the standard genetic optimization, and the improved ligand pose that is obtained from this local optimization is the input for the crossover portion of the algorithm. The original local search method is based on the Solis-Wets random optimization algorithm \cite{Solis1981}. 

\subsection{AutoDock-GPU with OpenCL}
General purpose graphics processing units (GPUs) are used to accelerate dense numerical calculations in a variety of settings, from high-performance computing (HPC) facilities to data centers and cloud resources. There are several commonly used application programming interfaces (APIs) that are used to program NVIDIA GPUs including CUDA, and OpenCL. A few docking programs have in the past five years made use of GPUs for accelerating their calculations \cite{Fang2016,McIntosh-Smith2015,Solis-Vasquez2017}. Recently Scripps Research, in collaboration with TU Darmstadt, developed an accelerated version of their AutoDock4 program using OpenCL, which provided up to 50$\times$  speedup over the single-threaded CPU version \cite{Solis-Vasquez2017,santos2019accelerating,Santos-Martins2019}. OpenCL was chosen as the programming model for using the accelerator as it provides code portability to various types of architectures, for instance GPUs, CPUs, and FPGAs, from multiple vendors. This program has been named AutoDock-GPU and while it uses the same physics-based empirical potential energy function, it includes new algorithmic additions and changes that can improve both performance and the quality of the results \cite{Solis-Vasquez2017,santos2019accelerating, Santos-Martins2019}. In particular, within the AutoDock-GPU program, there are now two possible algorithms for the local search that can be employed, (1) the Solis-Wets (SW) method of random optimization, and (2) the ADADELTA gradient-based method \cite{zeiler2012adadelta}. The ADADELTA method was added to improve the local refinement of results, especially for ligands with larger numbers of torsions. AutoDock-GPU uses similar run parameters as AutoDock4, with some minor changes, including a renaming for some parameters, changes in some defaults, and some different hyperparameters. The run parameters important to the current study include the \texttt{nrun} parameter, which designates the number of different independent complete LGA calculations that are performed in one instance of the executable. Each \texttt{nrun} results in a separate final pose, and is output in the output files. A drug screen would take the best scoring pose from these independent outputs. In addition, a recently developed feature is the ability to set the \texttt{autostop} parameter, which allows the local search algorithm to finish prematurely (with respect to the value set in the \texttt{nev} parameter that sets the number of total iterations of the LGA algorithm per run), if the energy value has not changed a sufficient amount, measured by the standard deviation over the past 10 iterations.
\vspace{-0.5em}
\subsection{The Summit supercomputer}
The Oak Ridge Leadership Facility (OLCF) at the Oak Ridge National Laboratory (ORNL) is an open science computing facility that supports HPC research. The OLCF houses the Summit supercomputer, an IBM AC922 system consisting of 4608 large nodes each with six NVIDIA Volta V100 GPUs and two POWER9 CPU sockets providing 42 usable cores per node. Currently there is no NVIDIA driver support for OpenCL on POWER9 architectures for GPU. Summit is harnessed by hundreds of research groups as part of its multiple open science allocation grants, including the INCITE, ALCC, and Director's Discretion awards and most recently, as part of the COVID-19 Consortium created for supporting computational research aimed at combating the COVID-19 pandemic (\url{https://www.xsede.org/covid19-hpc-consortium}). The ability to leverage the Summit supercomputer's GPUs to perform massive high-throughput docking screens for antivirals against the SARS-CoV-2 virus was recognized by our groups as an important resource that was not accessible for AutoDock-GPU due to this lack of OpenCL support. We therefore have worked to create a new version of this program using the CUDA API, and in addition, to design programmatic changes that facilitated optimized docking screens using large ligand datasets. Here we describe these ports, optimizations including latency hiding facilitated by overlapping set-up and calculation, and inclusion of a convenient input option for processing hundreds to thousands of files more efficiently.

\subsection{Related Work}
Several previous efforts have enabled the deployment of docking codes such as AutoDock4 and AutoDock Vina on large compute resources, from clusters to supercomputers. Vina MPI used an MPI wrapper to enable the simultaneous launching of thousands of parallel AutoDock Vina executables and was run on the OLCF Jaguar and Titan supercomputers \cite{Ellingson2013}. A similar effort at the Lawrence Livermore National Laboratory resulted in a program called VinaLC \cite{Zhang2013b}. Recently a GNU-Bash based pipeline was created for docking millions of ligands on clusters \cite{gorgulla2020open}. GPU acceleration has also been included in docking programs. GeauxDock created a novel energy function consisting of physics-based and statistical terms that ran on GPUs \cite{Fang2016}, but has seen limited application or updates. To our knowledge, the version of AutoDock-GPU described in this paper is the only docking program using the AutoDock4 potential that is optimized for the most recent GPUs and also adds the capability to optimally process thousands of ligand input files sequentially with a single executable. 

\subsection{COVID-19 and drug discovery for SARS-CoV-2 antivirals}
The ability to rapidly dock millions to billions of ligands to protein receptors to help find inhibitors for viral proteins could be invaluable to experimental drug discovery efforts. Correspondingly, harnessing leadership computing resources for drug discovery efforts against the SARS-COV-2 virus could provide an invaluable resource with which to battle the current COVID-19 pandemic \cite{Parks2020}. Applying this approach is part of the collaborative international research effort recently formed around discovering therapeutics to mitigate morbidity and mortality caused by the disease. The work presented here takes the first step in facilitating the deployment of a GPU-enabled, open source docking program on the Summit supercomputer, and optimization for high-throughput docking of large numbers of ligands in a single instance. With this capability in place, future work involving the deployment of efficient platforms to enable running the AutoDock-GPU program at scale on all of Summit's 27,648 GPUs can be pursued.

\section{Methods}
\begin{figure}
    \centering
    \includegraphics[width=\columnwidth]{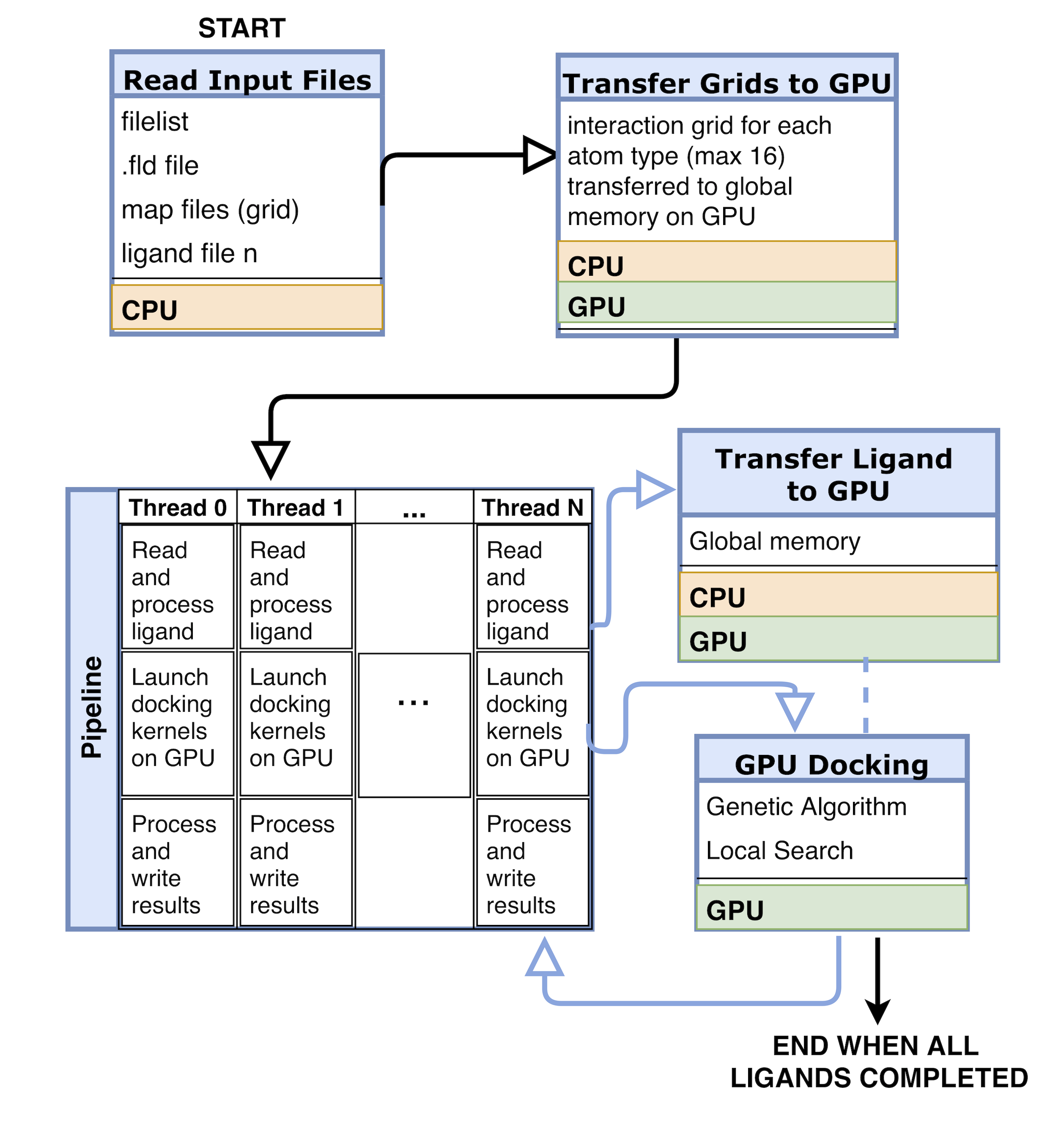}
    \caption{Schematic of new HPC-friendly AutoDock-GPU program, showing the OpenMP threading-based pipeline for hiding ligand input and staging, and the receptor-reuse functionality for docking many ligands to a single receptor. Noted are locations (CPU, GPU) where steps are executed. Black arrows indicate operations performed once, and blue arrow are those performed multiple times. }
    \label{fig:program}
%JVV I thought we weren't allowed to monkey with spaces?
\vspace{-1em}    
\end{figure}

A number of additions and changes were made to the AutoDock-GPU program to create a version that made effective use of the nodes on Summit and supported massive, high-throughput docking screens. In particular we focused on the case where a very large number of ligands (millions to billions) would be docked to a single receptor. Fig.~\ref{fig:program} illustrates the design of this version of the program, showing the addition of OpenMP threading for creating a pipeline that hides latency from file I/O by staging ligand files while the GPU is busy docking. In addition, the receptor data is now explicitly reused, further reducing I/O when docking thousands of compounds to the same protein target. These changes are detailed in the following subsections.

\subsection{Conversion of OpenCL version to CUDA and optimizations}

The original AutoDock-GPU program developed by Scripps and TU
Darmstadt was written in OpenCL to improve portability across GPU vendors, and enabled threaded execution on CPUs. However, OpenCL is not supported on all platforms, including the combination of NVIDIA GPUs and POWER9 CPUs present on Summit. Therefore it was necessary to port AutoDock-GPU to CUDA to leverage the V100 GPUs on Summit to run AutoDock-GPU at scale. This provides not only large speedups over CPU-based code but also a reduction in power consumption for the same compute task on Summit \cite{gharaibeh2013energy}. This porting is not effortless, due to philosophical and design differences between OpenCL and CUDA. OpenCL was created in 2009 as an open and portable alternative to CUDA \cite{Stone2010}, and has maintained the commitment to remain free of hardware-specific programming elements. While the OpenCL API resembles CUDA's API in many ways, it is a vendor-neutral solution with a focus on portability across CPUs, GPUs, FPGAs, and embedded devices, and thus cannot expose architecture-specific hardware features. CUDA, on the other hand, has been developed in tandem with NVIDIA GPUs hardware and exposes new architectural features with each release. While OpenCL has the advantage of running on any device that supports it, porting OpenCL code to CUDA also provides significant opportunities to exploit those vendor-specific hardware advances, which would otherwise squander the transistors on NVIDIA GPUs that enable them. 

In order to create an initial port that could reproduce the OpenCL results, we transcribed the AutoDock OpenCL version (forked April 3rd from the ``develop" branch) by hand to CUDA. This created an initial executable which performed at roughly half the speed of the OpenCL version. Next, we applied basic optimization techniques to the code like replacing calls to \texttt{pow} with intrinsics, and dynamically allocating shared memory to allow more threads per processor. These first optimizations enabled the CUDA version to run 2.8$\times$ faster than the OpenCL version on an NVIDIA RTX2080TI GPU for the ADADELTA algorithm. Back-porting these optimizations to the OpenCL code showed little to no benefit there, for the following reasons: OpenCL already has intrinsics for \texttt{pow(float, int)}, and the shared memory amount is known to OpenCL at compile time. 

\subsection{Optimizing for thousands of consecutive docking calculations}
Although the OpenCL and CUDA implementations of AutoDock-GPU were able to accelerate docking calculations significantly, AutoDock-GPU was designed to take a single receptor and a single ligand as command-line inputs. If one wished to dock multiple ligands with a single receptor (or vice versa), one would have to run the executable separately for each ``job.'' For a large ligand set, this is both tedious for the user and inefficient in terms of resources. Here we describe several steps taken to maximize performance in the context of massive docking calculations.

\subsubsection{A multiple-files option and file reuse}

To improve the code for this use case, we enabled the user to provide a list of protein and ligand files, accessible through a \texttt{-filelist} flag. The program then loops over the provided list and performs the docking calculations for each. To further optimize the performance of GPU-based high-throughput docking of thousands of ligands to a single receptor, we enabled the program to reuse the 16 receptor maps of up to hundreds of megabytes containing the interaction grid information for atoms in ligands rather than re-read the map files for every new ligand.

\subsubsection{OpenMP threading for overlapping set-up}
 
After enabling multiple jobs to run consecutively in the same executable, we could begin performance optimization of this new pipeline. Depending on the parameters specified, a significant portion of the run time is spent in the setup and post-processing phases. These phases are performed on the host and include I/O, allocations, and some initial and final calculations less suitable for GPU. If jobs are executed serially, the GPU is idle during these phases and thus underutilized. To reduce this latency, we introduced OpenMP directives so that CPU-based threads could work in parallel to load ligand input files, transfer them onto the GPU, launch the GPU-based docking kernels, and process and write the output of the calculation. This is illustrated in Fig.~\ref{fig:program}. To ensure that jobs do not conflict, all CPU-GPU communication (besides the initial receptor map setup described above) and the docking algorithm itself (consisting of multiple CUDA kernels) are placed in a function called from an \texttt{omp critical} section. As a result, the OpenMP threads perform the setup and post-processing phases of each job in parallel, while queuing up to use the GPU one by one for the docking simulation itself. Combined with the reduction in setup time from the optimizations detailed below, this threading approach is sufficient to hide almost all setup and post-processing time and ensure the GPU is rarely idle.

\subsubsection{Reuse of GPU context and of receptor grids on GPU}

We also improved handling of GPU memory and host-device communication. Instead of creating and destroying the CUDA context anew for each ligand input, the context is now created once and GPU memory is preallocated and reused for each ligand. Similarly, rather than transfer the receptor grid maps onto the GPU for every ligand, all of the receptor's grids are sent once during initialization and left on device. Because each ligand requires a different subset of these grids, a mapping was required to point to the grids necessary for a specific ligand docking. Historically, creation of CUDA contexts in memory, GPU memory allocation, and transfer of data to GPU have been bottlenecks in GPU programming. On Summit, with the improved NVLINK2 interconnect, data transfer onto the GPU is less expensive, but still considerable. The V100 GPUs on Summit contain 16 GB of global memory, which is enough to store all possible receptor interaction grids for supported atom types.

\subsection{Performance and validation testing on OLCF resources: Summit and DGX-2}
We tested the performance of the new CUDA version of AutoDock-GPU with and without the OpenMP-based pipeline and compared to performance of the OpenCL version using a NVIDIA DGX-2 appliance hosted at ORNL. The DGX-2 contains 16 NVIDIA V100 GPUs and dual Intel Xeon Platinum 8168 CPUs containing 24 cores. The x86 architecture for the DGX-2 platform permits us to make direct comparison between the original OpenCL AutoDock-GPU implementation and the CUDA port. Both executables were built within an Ubuntu 18.04 Singularity container with CUDA 10.1 and version 7.5.0 of the GNU compiler collection, which implement OpenCL 1.2. These tests were performed on a set of 42 experimental crystallographic structures of ligands bound to enzyme active sites, obtained from the Research Collaboratory for Structural Bioinformatics (RCSB) Protein Data Bank (PDB) \cite{berman2000protein}, and converted to input files for AutoDock-GPU using OpenBabel program \cite{OBoyle2011}. The set of inputs and the PDB IDs for these structures can be found at \url{https://github.com/diogomart/AD-GPU_set_of_42}. This set is hereafter referred to as S42. The S42 dataset contains a variety of different ligand sizes and numbers of torsions, or rotatable bonds, and a range of different sized search boxes defined by the input grid coordinates.

We then tested the performance of the CUDA/pipeline version of AutoDock-GPU on a large benchmark set of ligands on Summit. For Summit tests we used GNU compiler collection version 6.4 and CUDA version 10.1.243. Summit currently runs the Red Hat Enterprise Linux Server version 7.6 (Maipo). The ligand dataset consisted of 9,000 ligands taken from the full SWEETLEAD database \cite{Novick2013} (as acquired in March of 2020), but with ligands containing atom types unsupported by the AutoDock Utilities tools \cite{Morris2009,Solis-Vasquez2017} removed, and supplemented with ligands from the NCI. This data set will be referred to as SN9000. Performance was measured using both local search algorithms (Solis-Wets and ADADELTA), and compared to the performance of both AutoDock4 and AutoDock Vina on Summit. The SARS-CoV-2 endoribonuclease protein (NendoU) crystal structure recently deposited on the RCSB PDB, PDB-ID 6VWW \cite{Kim2020}, was the target receptor. 

Finally, we explored in detail the different optimizations which were added, namely, the application of the OpenMP pipeline, the addition of context reuse, and the reuse of the receptor grids by storing all of them on the GPU's global memory. We performed these tests on Summit using a single resource set on a single compute node which consists of a GPU, 1 (for non-pipelined version) and 7 (for pipelined version) physical CPU cores with one thread per core with the pipeline using 1 and 7 OpenMP threads. For this test we picked a random subset of about 400 ligands from the SN9000 ligand set, and also docked to the NendoU receptor with a search box of size 22.5 \AA~ per side using both local search methods (Solis-Wets and ADADELTA). We chose ligands with small numbers of torsions ($\leq 10$) to test the performance of the Solis-Wets method, and ligands with more than 10 torsions for ADADELTA. We performed the test using the number of runs (\texttt{nrun}) parameter set to 10 and 100 for both cases, and with the newly implemented autostop functionality both on and off. For all receptors, any bound small molecule ligand, ions, and water molecules were removed from the structures using MOE \cite{Vilar2008}. Hydrogen atoms were then added, and each structure was saved in its apo form. The AutoDockTools script (\texttt{prepare\_receptor4.py}) was then applied to convert the PDB files to PDBQT format, and AutoGrid4 was employed to generate the atom-specific affinities, electrostatic potential, and desolvation potential maps for each receptor. Details of experiments on viral Mpro (section 4) are given in that section. Molecular images were made with VMD \cite{Humphrey1996} and UCSF Chimera \cite{pettersen2004ucsf}.

\section{Results: validation and performance evaluation }
Here we describe the results of performance testing and validation of the CUDA/pipeline version of AutoDock-GPU on several potential use-cases for high-throughput docking. The parameters that are varied in these different situations include the number of ligands docked against a receptor, the size of the allowed search box, the number of torsions in the ligands, the local search algorithm used, and the number of replicas of the calculation that are performed for each docking (the \texttt{nrun} parameter). We break down the performance gains imparted by the different improvements detailed in the Methods section. We also confirmed that the results obtained with the new version are consistent with those obtained using the original OpenCL version.

\subsection{Comparison of OpenCL and CUDA version and effect of pipeline}
\begin{figure}
    \centering
    \includegraphics[width=0.8\columnwidth]{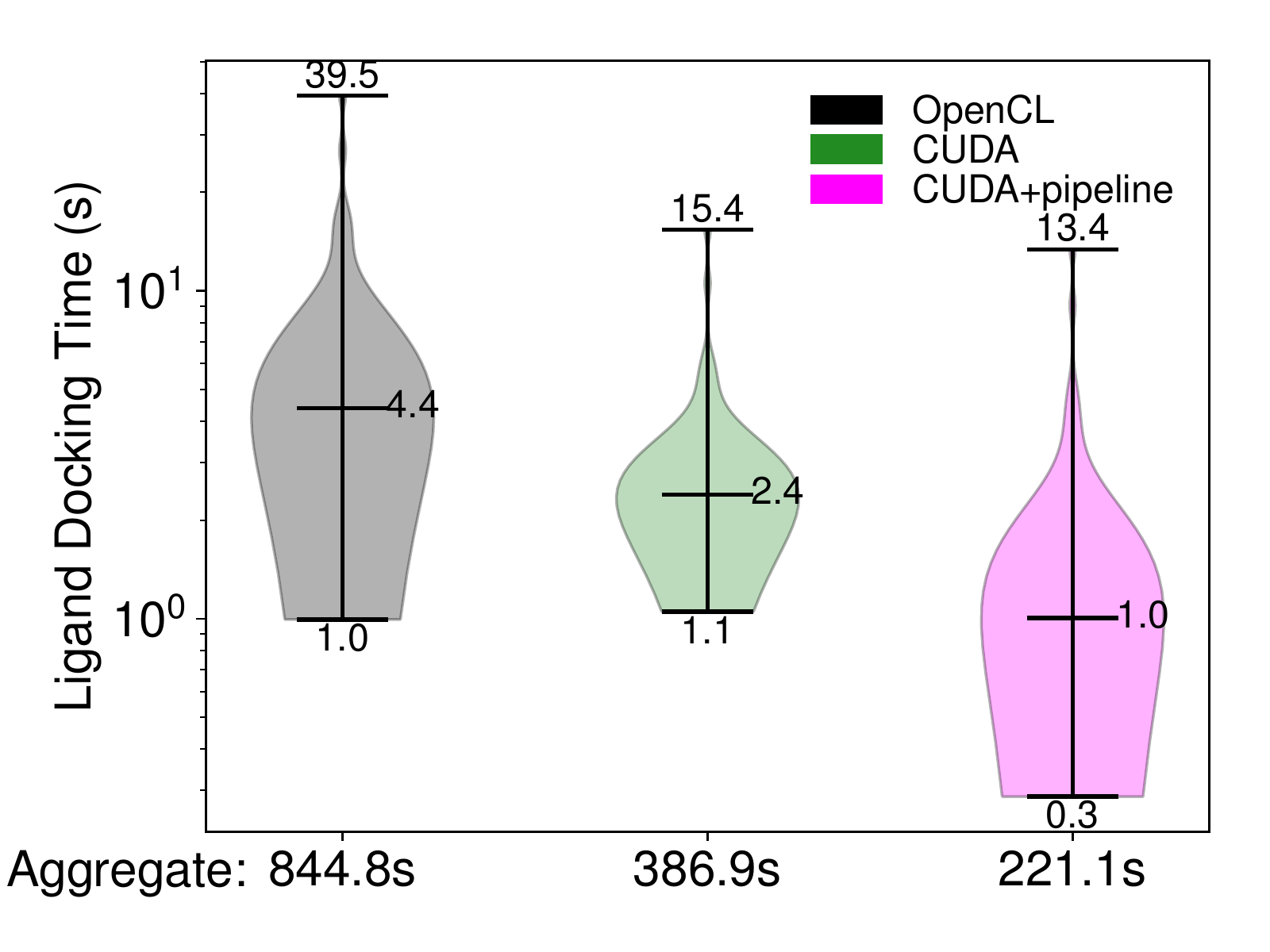}
    \caption{AutoDock-GPU performance, using the ADADELTA algorithm, on a DGX-2 on a set of 42 receptors with 3 ligands each with \texttt{nrun=10}, permitting the OpenCL, CUDA, and pipelining features to be compared directly.}
    \label{fig:openclcomparison}
\end{figure}

Results for running the S42 benchmark on the DGX-2 are shown in Fig.~\ref{fig:openclcomparison}. Each of the 42 receptor-ligand pairs was tested with three random starting conformations of the corresponding ligand. The tests were performed using \texttt{nrun} set to 10 and using the ADADELTA local search method, with all other parameters set as default. The CUDA version here shows a 1.8$\times$ mean speedup over the OpenCL version with this dataset, and the pipeline provides an additional 2.4$\times$ speedup over the un-pipelined CUDA version, resulting in an overall 4.4$\times$ mean speedup over the original OpenCL version, for this particular set of receptors, ligands, and box sizes, and these run parameters.

\subsubsection{Re-docking validation results}

\begin{figure}
    \centering
    \includegraphics[width=0.8\columnwidth]{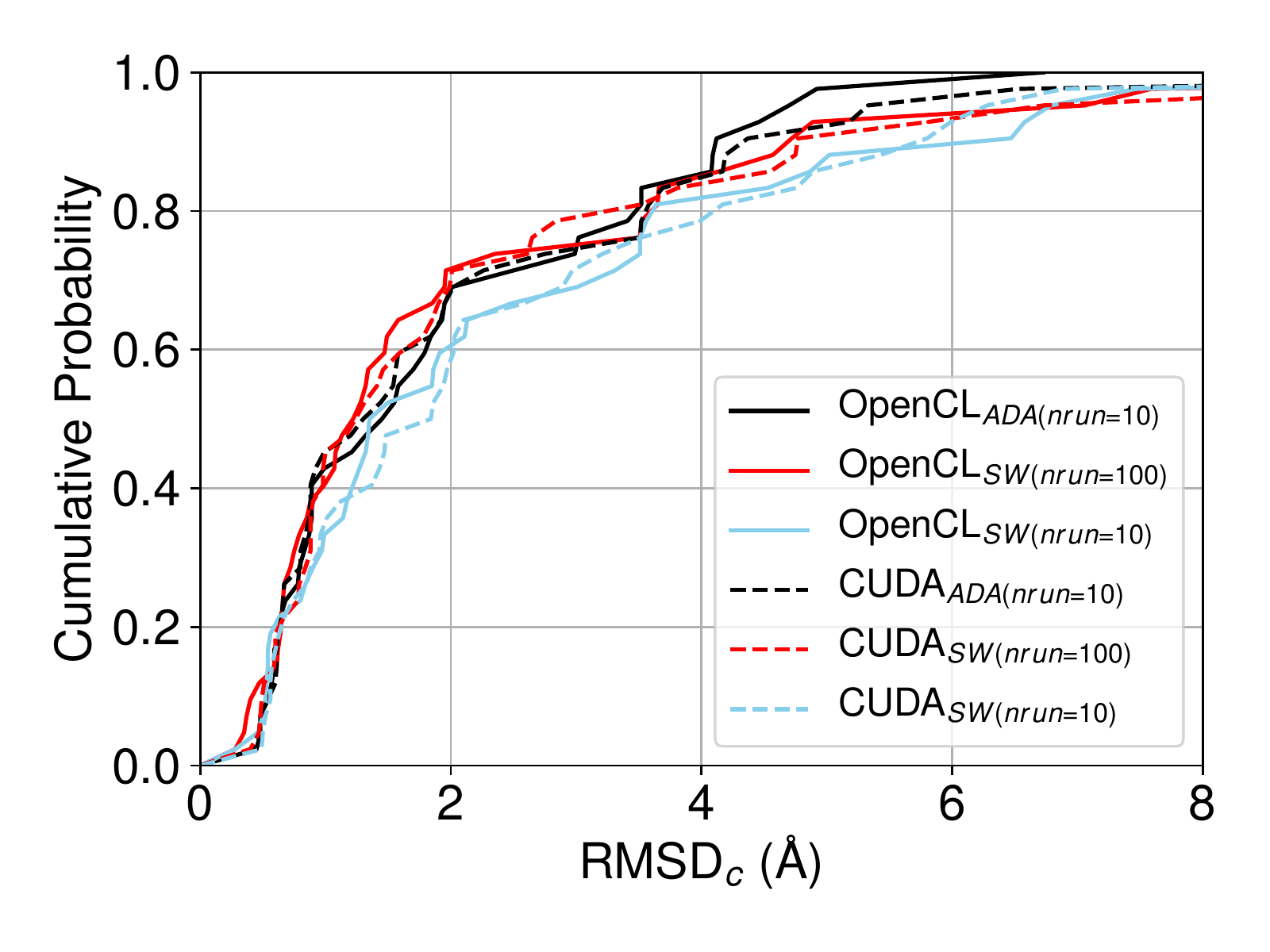}
    \caption{Cumulative distribution for RMSD for the redocked S42 set against the initial crystal structure, noted here as RMSD$_c$. The cumulative distributions were calculated for both OpenCL (solid) and CUDA (dashed) implementations and both ADADELTA or Solis-Wets local searches.}
    \label{fig:nrunsrmsd}
\end{figure}

Another metric over which to compare docking algorithms are the generated poses themselves. Leveraging the experimentally determined positions for ligands, we can check how well the predicted bound poses match experiment. This is known as re-docking. Because the S42 dataset consists of experimental structures of ligands bound to receptors, it can be harnessed to not only validate the ability of the AutoDock-GPU program to reproduce experimental ligand poses, but to also validate the consistency of outputs after major changes to the program. Here we show the cumulative root mean squared (Euclidean) distance (RMSD) distribution between the final best pose from docking and the crystallographic pose from the S42 set (Fig.~\ref{fig:nrunsrmsd}). The RMSD is a standard measure of the three-dimensional similarity between conformations of a molecule. Note that some ligands within this test set are intentionally difficult, with a large number of torsions and a commensurately large search space, and so not all poses found are near the experimental position. Despite these challenges, the median RMSD over the full 42 ligand set ranges from 1.28 to 1.85\,\AA~for the CUDA implementation of AutoDock-GPU, depending on the local search algorithm. This indicates useful agreement with the experimental poses. As expected, the docking quality improves with increased computational effort, although the improvement is not linear, and usable results can be obtained with a comparatively small number of runs, particularly with the ADADELTA local search algorithm. The small differences in the results between OpenCL and CUDA implementations within Fig.~\ref{fig:nrunsrmsd} are not indicative of qualitatively different results, are instead consistent with the stochasticity inherent in docking algorithms.

\subsection{Influence of size of search space}

\begin{figure}
    \centering
    \includegraphics[width=0.8\linewidth]{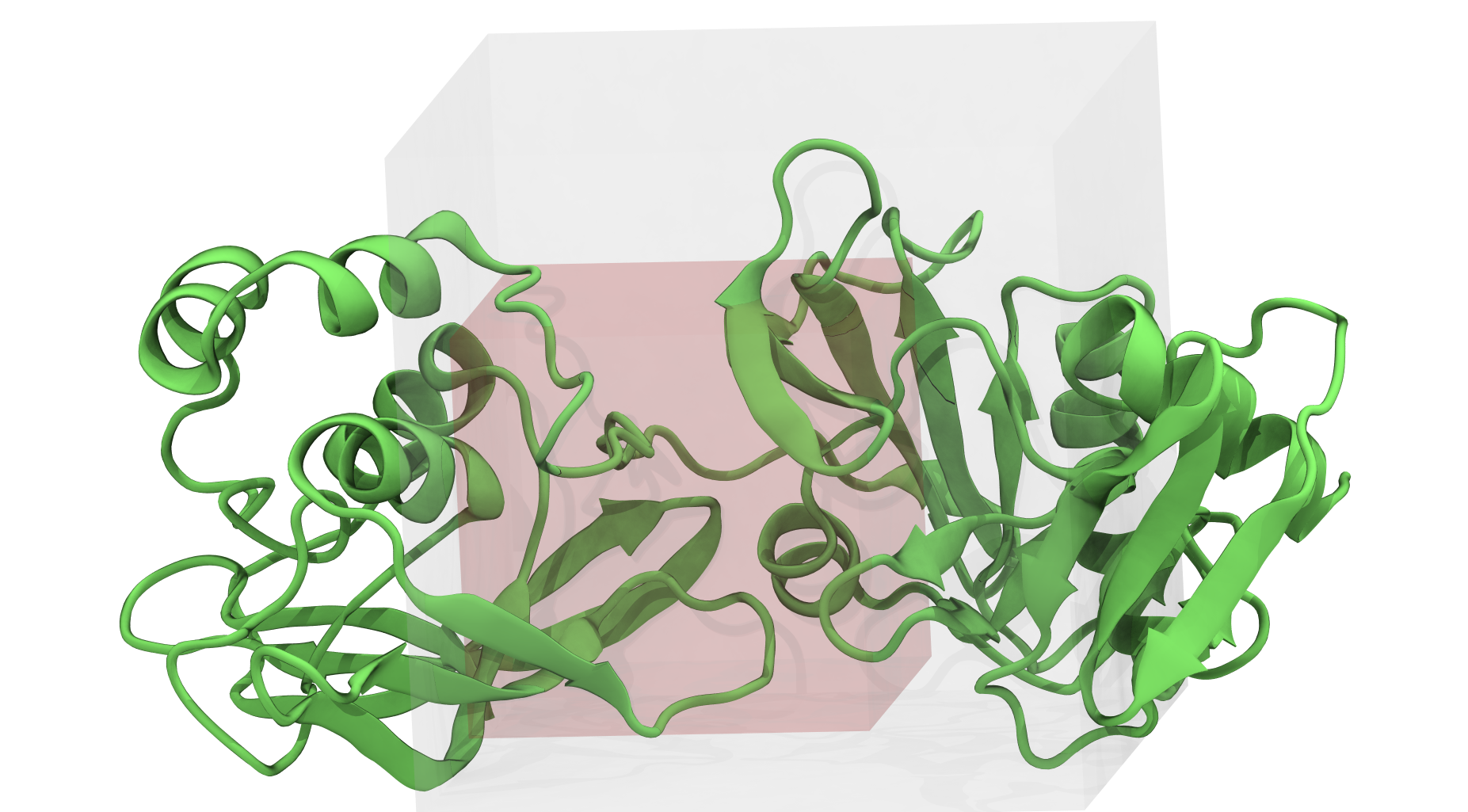}
    \caption{Blind docking (large box, light grey) and focused docking (small box, red) regions around the SARS-CoV-2 (green cartoon) endoribonuclease protein active site (PDB-ID 6VWW). Blind docking uses a larger search space (gray), and the small box (red) focuses on a small region directly around the active site.}
    \label{fig:blindtargeteddocking}
\end{figure}

For these tests we used the SARS-CoV-2 endoribonuclease protein (NendoU) crystal structure recently deposited on the RCSB PDB, PDB-ID 6VWW \cite{Kim2020}. In addition, we tested both the Solis-Wets local search method, and the ADADELTA method introduced with AutoDock-GPU.
We created two test cases for evaluating two extremes of search-space size. The first is an exhaustive search over the protein, known as ``blind'' docking, together with a complete set of ligand torsional degrees of freedom. The second uses a focused, small search box, limited to the active site region that must be pre-determined together with a subset of the ligand dataset that contained only those ligands with $\leq 10$ torsions. Fig.~\ref{fig:blindtargeteddocking} shows the different search spaces on the receptor.

\subsubsection{Blind docking with many degrees of freedom}

\begin{figure}
    \centering
    \includegraphics[width=0.8\linewidth]{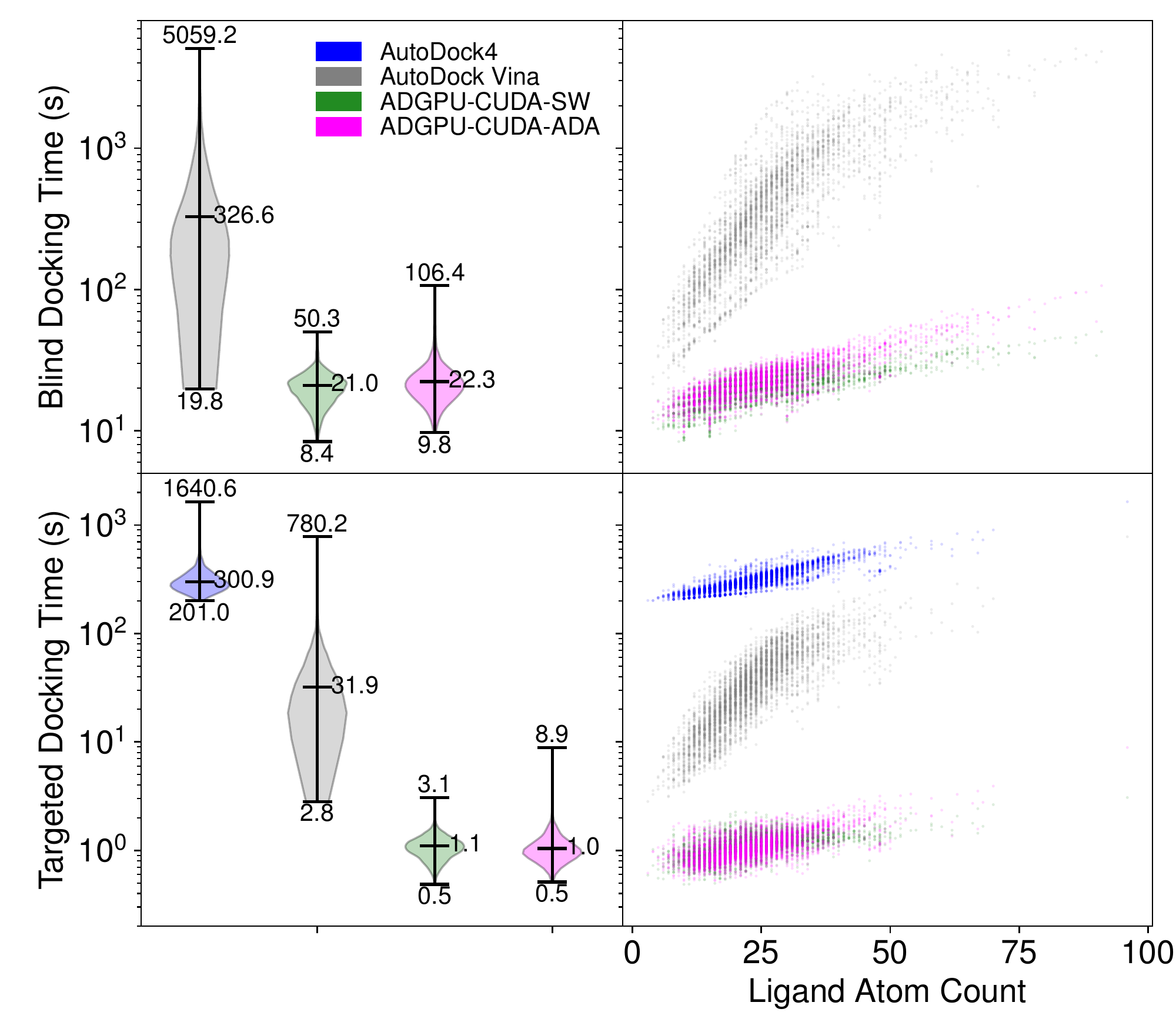}
    \caption{Runtime distribution for docking calculations under varying conditions with alternative docking programs. Left: violin plots of runtime distributions with maximum, median, and minimum runtimes for ligands indicated adjacent to their position within the distribution. Right: runtime distributions illustrating growth in runtime as ligands increase in size.}
    \label{fig:runtimecomparison}
    \vspace{-1.2em}    
\end{figure}

In the large search space test the ligands were allowed to bind anywhere in a large search volume (gray region, Fig.~\ref{fig:blindtargeteddocking}) which was a box 40 \AA~ per side centered on the active site. In addition, we used the full set of ligands which included those with $\geq 10$ torsional degrees of freedom, or rotatable bonds. We set (\texttt{nrun}=100) in order to explore this larger search space sufficiently. The dockings were performed with both the SW and the ADA-DELTA methods.

\subsubsection{Small Search Box with Fewer Degrees of Freedom}
 For the small search space test, we docked to the known binding site. In this case, only a small region (22.5 \AA~ per side) of the protein was explored for ligand binding (red box, Fig.~\ref{fig:blindtargeteddocking}), selected based on the final binding location for the majority of ligands in the blind docking trials, and only those ligands with $\leq 10$ torsions. The smaller search box coupled to selecting for simpler ligands in this test means that fewer runs were needed to saturate the space and we therefore set (\texttt{nrun}=10). 

\subsubsection{Search Space Tests and Comparison to Other Docking Programs} 
The two previous test cases were also used to compare the performance of AutoDock-GPU to equivalent calculations using the CPU-only AutoDock4 \cite{Morris2009} or AutoDock Vina \cite{Trott2009} programs when performed on Summit. For the large search space test, we compared time to solution using AutoDock Vina against the time to solution for AutoDock-GPU on Summit. AutoDock4 was excluded from this analysis, as the high number of runs required to sample the space (\texttt{nrun}=100) would routinely exceed the allowed walltime on Summit. For the small search space test, a lower \texttt{nrun} setting permitted AutoDock4 to be added to the comparison set. The results of these runtime tests under these two conditions are summarized in Fig.~\ref{fig:runtimecomparison}. For the large space test, AutoDock Vina runtimes are observed to grow rapidly as a function of ligand complexity. As a consequence, while the median docking times are only a factor of 16 slower in the large space case, the aggregate run times for all AutoDock Vina calculations are a factor of 25 slower due to particularly poor performance for the largest ligands.
By contrast, AutoDock-GPU tackles even these challenging ligands in under two minutes (Fig.~\ref{fig:runtimecomparison}). The general runtime distribution for AutoDock4 and AutoDock-GPU on Summit is similar, with additional ligand complexity only modestly increasing the runtime for an individual ligand (Fig.~\ref{fig:runtimecomparison}).
However, whereas AutoDock4 has significantly longer runtimes even for the simple ligand set selected, AutoDock-GPU leverages the parallelism inherent to the GPUs to bring the runtimes down significantly. The large reduction in runtimes and narrower time distribution may permit large ligand sets to be screened on Summit on a routine basis.

\subsection{Performance improvement from individual components of new high-throughput design on Summit} 
\begin{table}
  \caption{Performance Improvements from Sequential Optimizations: Solis-Wets. Shown is mean time in seconds over ligands run sequentially on one GPU using the pipeline with with either 1 or 7 threads. Ligands in this set contained 10 or fewer torsions. CR: context reuse; F/RR: receptor file and receptor grid reuse on GPU. Each optimization is added on top of the previous ones going from left to right.}
  \label{tab:perf_sw}
  \begin{tabular}{cccccl}
    \toprule
    \textbf{\texttt{nrun}}& \textbf{1 thread} &\textbf{7 threads} &\textbf{CR}&\textbf{F/RR}\\
    \midrule
    \textbf{10}&4.7&1.4&1.4&0.5\\
    \textbf{100}&8.8&4.6&2.1&2.1\\
\bottomrule
\end{tabular}
\vspace{-.5em}
\end{table}

We also performed a systematic test of the incremental contributions of each of the optimizations to the overall speedup of the new HPC-centric version of AutoDock-GPU. 412 randomly selected ligands from the SN9000 dataset were picked for this analysis. We computed the average time in seconds per ligand for values of the \texttt{nrun} parameter set to 10 and 100, for the program with and without threading (by setting the \texttt{OMP\_NUM\_THREADS} environment variable to 1 and 7 respectively), and tested different stages of optimization the program, added incrementally: with addition of CUDA context reuse, and with the reuse of the receptor input files and of this data by making employing of GPU global memory to store all required grids for all ligands in a batch. Tables \ref{tab:perf_sw} and \ref{tab:perf_ad} show these effects for the SW and ADADELTA algorithms, respectively. Each new optimization is added on top of the previous one in these tables.
\begin{table}[h]
  \caption{Performance Improvements from Sequential Optimizations: ADADELTA. Mean  run time (in seconds) calculated as in Table \ref{tab:perf_sw}. CR: context reuse; F/RR: receptor file and receptor grid reuse on GPU. Each optimization is added on top of the previous ones going from left to right. }
  \label{tab:perf_ad}
  \begin{tabular}{ccccccl}
    \toprule
    \textbf{torsions}&\textbf{\texttt{nrun}}&\textbf{1 thread}& \textbf{7 threads}&\textbf{CR}&\textbf{F/RR}\\
    \midrule
    $\mathbf{\leq 10}$& \textbf{10}&4.8&1.4&1.4&0.7\\
    $\mathbf{\leq 10}$& \textbf{100}&9.7&5.5&4.3&4.4\\
    \textbf{ > 10}& \textbf{10}&7.5 &3.5&3.3&3.4\\
    \textbf{ > 10}& \textbf{100}&25.8 &23.9&20.4&20.5\\
\bottomrule
\end{tabular}
\end{table}

\begin{figure}
     \centering
     \includegraphics[width=0.6\columnwidth]{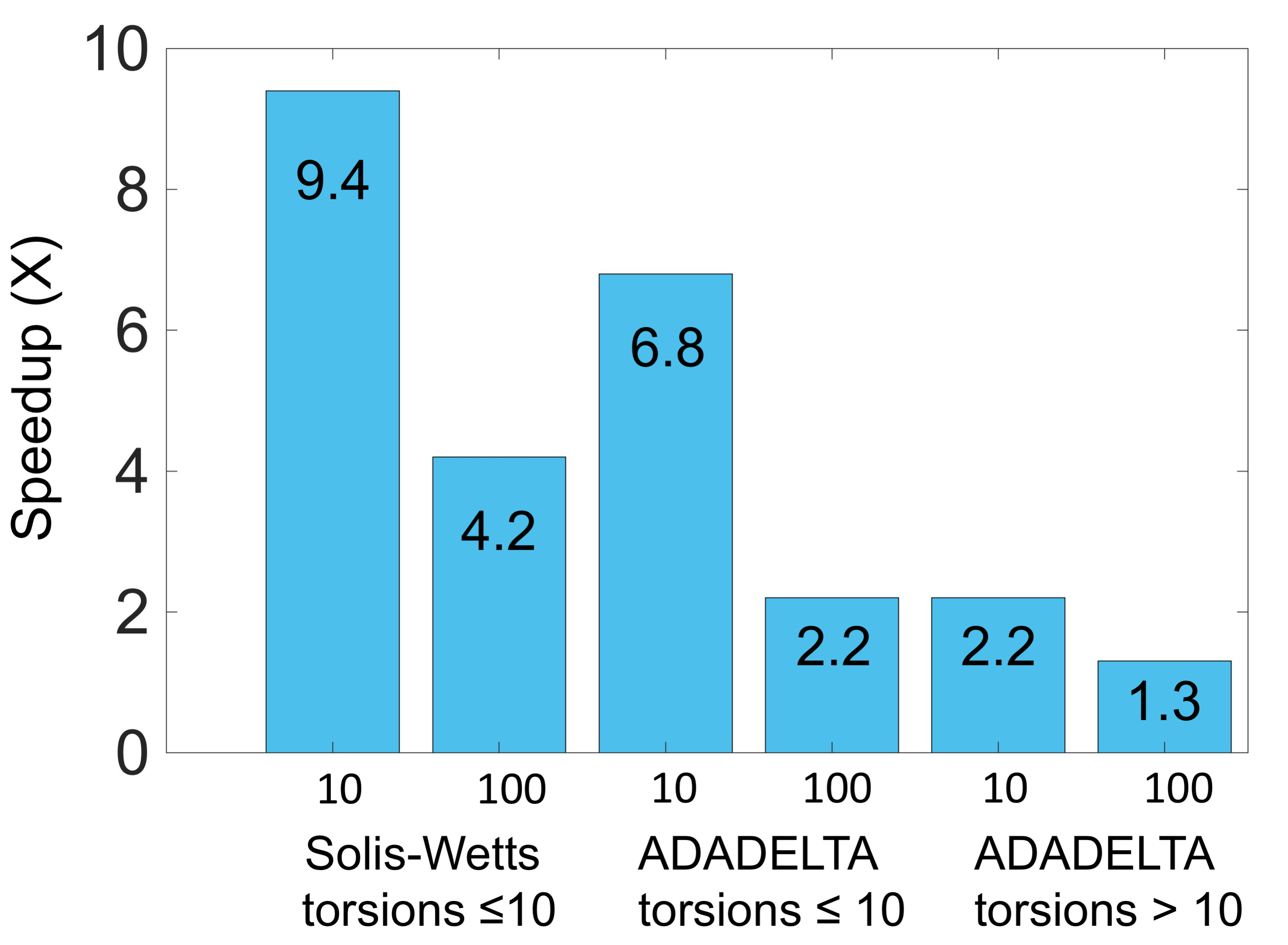}
     \caption{Total speedup over all optimizations for variations of local search algorithm, torsion number and \texttt{nrun} parameter (10 or 100) indicated by the numbers under the bars.}
    \label{fig:perf_all}
    \vspace{-0.5em}
 \end{figure}

For simple ligands under 10 torsions, both for SW and ADADELTA, the use of 7 threads with the OpenMP pipeline provided about 3 to 3.4$\times$ speedup for \texttt{nrun} = 10, but slightly less than a 2$\times$ speedup for \texttt{nrun} = 100. With \texttt{nrun} = 10, for both low torsion sets, the file and receptor reuse optimization provided a significant further speedup of about 2 to 2.8$\times$, while CUDA context reuse did not provide any additional speedup, while context reuse provided 2.2$\times$ speedup for SW with \texttt{nrun} = 100, and file and receptor reuse did not add performance. Fig.~\ref{fig:perf_all} summarizes the total speedup, with all optimizations, for all 6 cases.

The \texttt{autostop} parameter was added recently to the development version of AutoDock-GPU (OpenCL) and ported to the CUDA version. It allows the program to stop the local search if no progress is being made. Table \ref{tab:as_perf} shows the speedup provided by using \texttt{autostop}, shown without the pipeline (threads set to 1). This feature helps most for larger \texttt{nrun} values combined with smaller molecules.

\begin{table}
  \caption{Contribution of \texttt{autostop} feature to time to solution. Right two columns show mean run time per ligand in seconds. AS: \texttt{autostop}; alg, algorithm; SW, Solis-Wets; AD, ADADELTA. Threads used:1.}
  \label{tab:as_perf}
  \begin{tabular}{cccccl}
    \toprule
    \textbf{alg}&\textbf{torsions}&\textbf{\texttt{nrun}}& \textbf{AS off}&\textbf{AS on}\\
    \midrule
    SW&$\mathbf{\leq 10}$& \textbf{10}&4.7&4.5\\
    SW&$\mathbf{\leq 10}$& \textbf{100}&8.8&6.4\\
    AD&$\mathbf{\leq 10}$& \textbf{10}&4.8&4.7\\
    AD&$\mathbf{\leq 10}$& \textbf{100}&9.7&8.8\\
    AD&\textbf{ > 10}&  \textbf{10}& 7.5& 7.4\\
    AD&\textbf{ > 10}&  \textbf{100}& 25.8 & 25.6\\
\bottomrule
\end{tabular}
\vspace{-1em}
\end{table}

\section{Results: Experiments on SARS-CoV-2 main protease (Mpro)}

\begin{figure}
     \centering
     \includegraphics[width=0.7\linewidth]{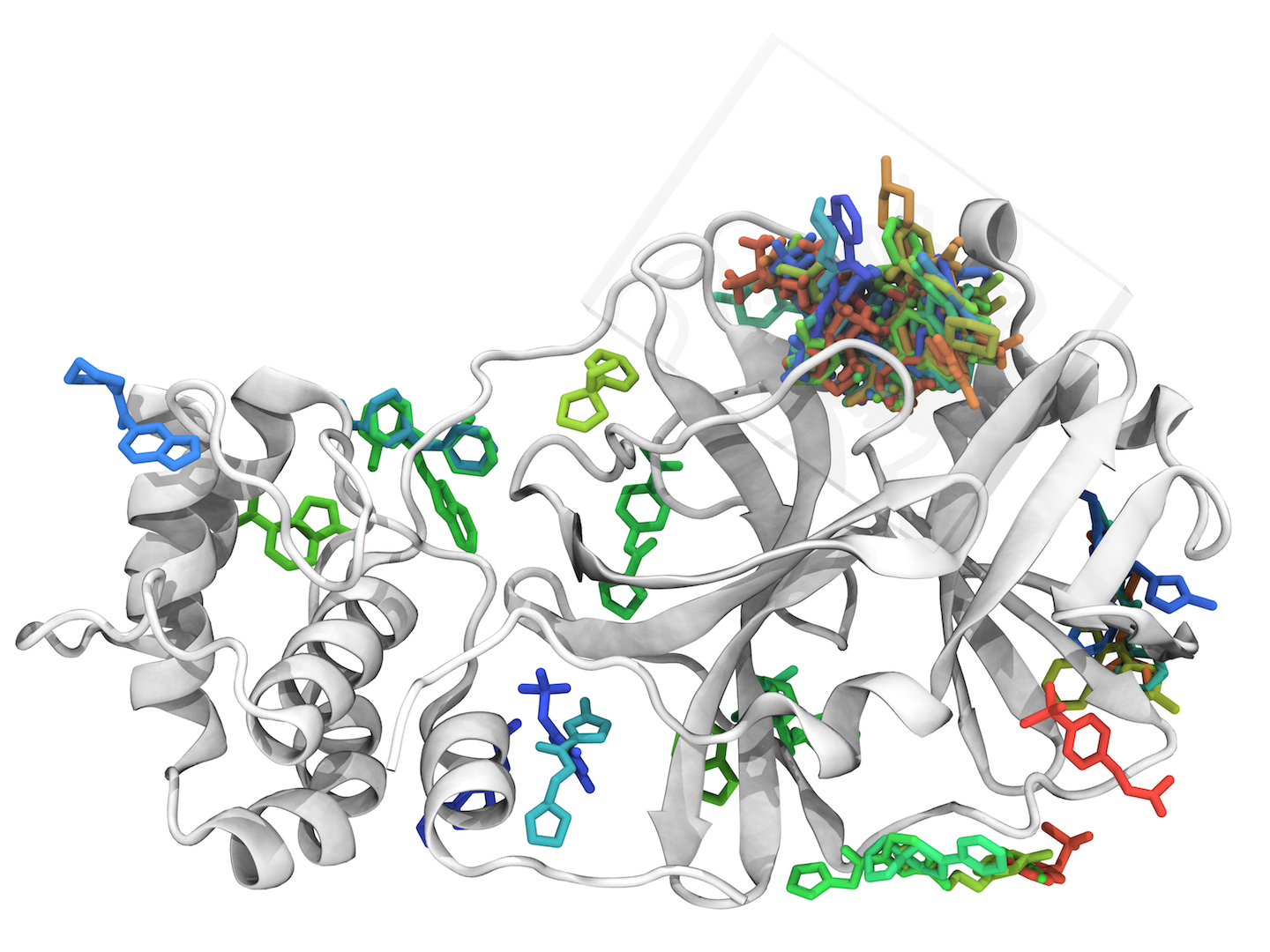}
     \caption{Ligand binding locations (multicolored sticks) observed in crystal structures of the SARS-CoV-2 main protease (white cartoon).}
    \label{fig:pdbdocking}
    \vspace{-1em}
 \end{figure}

 Many groups from around the world have been studying different SARS-CoV-2 proteins and solving their structure. The main protease (Mpro) in particular has attracted much attention as a potential drug target, and as such over 90 crystallographic structures containing a bound ligand have been released on the RCSB PDB, with many bound to the protease's active site region, and others bound elsewhere on the protein. We therefore performed some preliminary docking calculations on ligands taken from this set. These calculations allow us to examine calculated binding energies and docking poses. 
 Fig.~\ref{fig:pdbdocking} shows an overlay of some of these ligands bound to Mpro.

 From this set, 68 ligands are observed to interact with the enzyme active site, of which 22 are non-covalent ligands that have been crystallized. The remaining 46 ligands bound to the active site are covalently bound to the receptor, and this interaction cannot be represented in AutoDock Vina or AutoDock-GPU.
 Fig.~\ref{fig:pdbdocking} indicates that 25 crystallographically determined ligands bind non-covalently to Mpro, well outside of the active site region. All of these putative ligands are fragments, and therefore are relatively small, with no more than 4 torsions. 
 
\begin{figure}
    \centering
    \includegraphics[width=0.9\columnwidth]{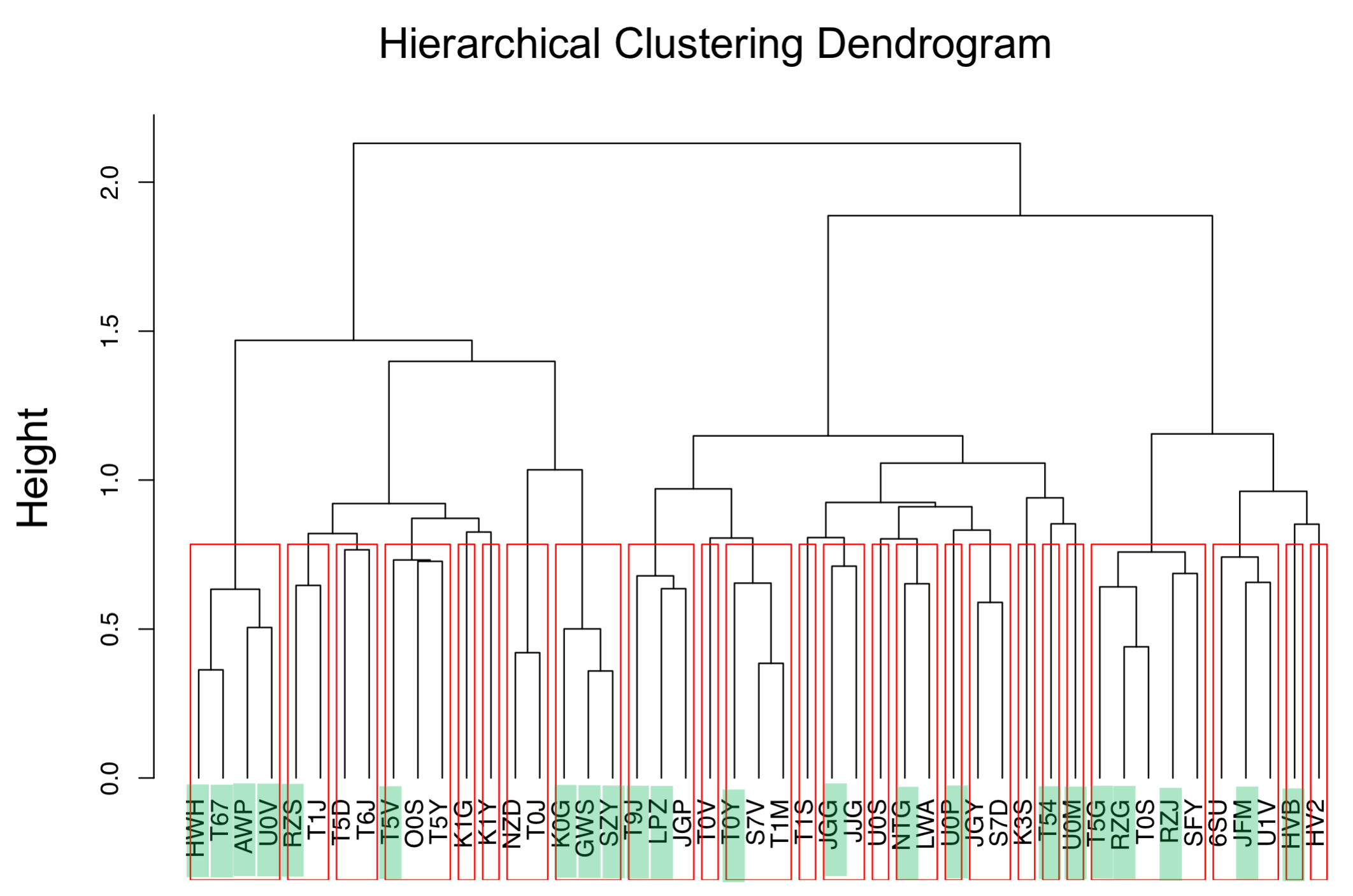}
    \caption{Dendrogram of hierarchical clustering using MDL keysets. The small molecules were divided into 24 clusters (clades) based on structural similarity. Active-site binders are highlighted in green.}
    \label{fig:cluster}
\end{figure}

 A hierarchical chemical clustering over all non-covalent binders to the protease was performed in chemical fingerprint space using MDL keysets \cite{durant2002reoptimization}, the Tanimoto similarity coefficient between compounds, and the Ward clustering linkage method with a clustering threshold of 0.8. The small molecules were divided into 24 clusters (clades) based on structural similarity (Fig.~\ref{fig:cluster}). All molecules in this set, both bound to the active site and outside of it, were found in similar clusters.
 Furthermore, many of the external binders were in the same clades as active site binders. It is possible, considering their closeness in chemical fingerprint space as shown by the clustering analysis, that the external binders also occupied the Mpro active site for some time before being displaced during crystallization.
 
\begin{figure}[h]
    \centering
    \includegraphics[width=0.7\columnwidth]{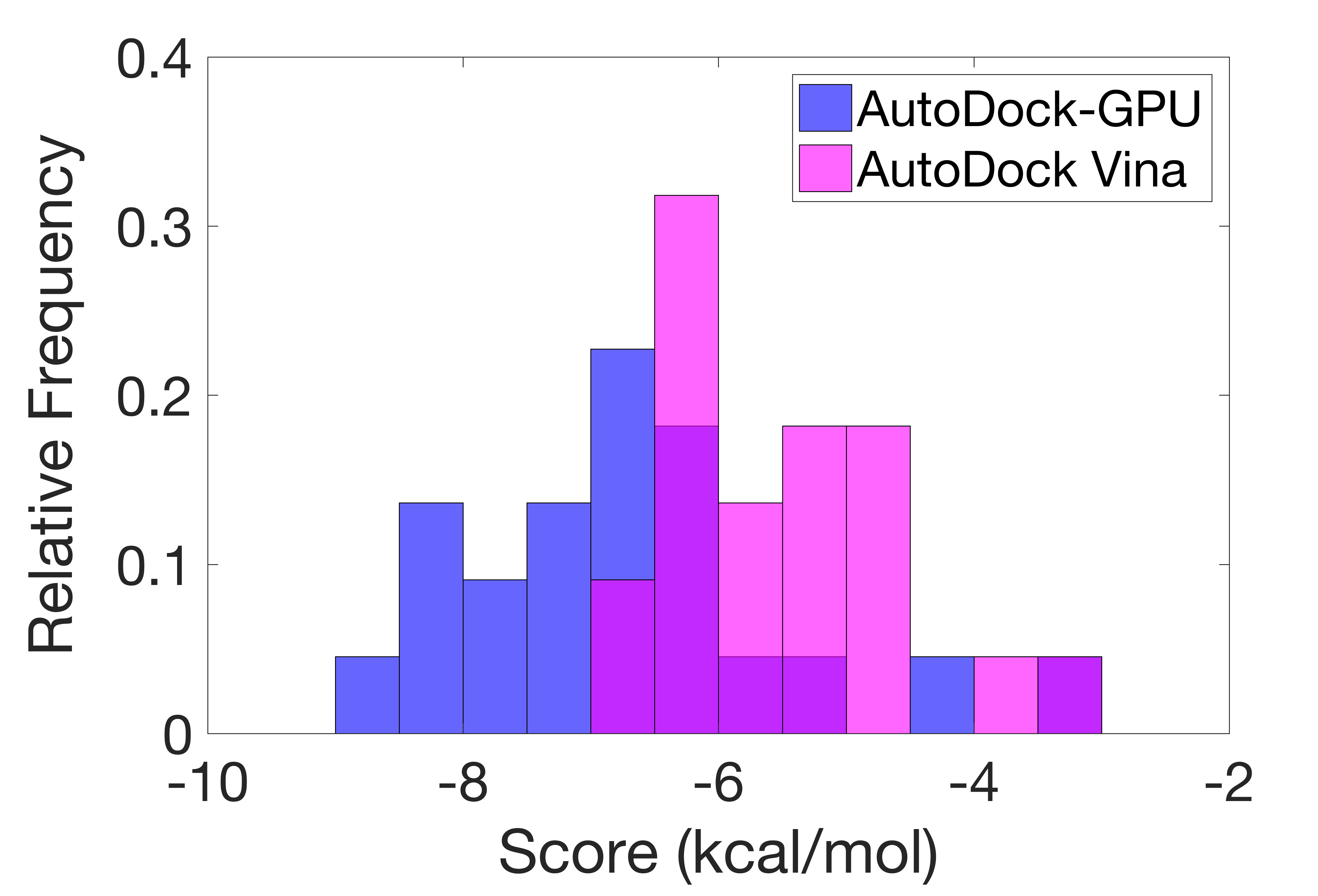}
    \caption{Histogram of the best binding free energies (scores) of ligands which were experimentally found to be bound non-covalently to Mpro active site, docked to Mpro structure PDB ID 5RE9 using AutoDock-GPU and AutoDock Vina.}
    \label{fig:hist_vin_ad}
\end{figure}

We docked both the 22 active-site binders and the 25 external binders to the Mpro receptor. For these tests all docking calculations were performed with SW with \texttt{nruns} = 100.  Interestingly, scores for both sets (active site and external binders) had a mean between -6.4 and -7.4 kcal/mol for all of the Mpro crystal structures tested, with best scores over each ligand set extending below -8.5 kcal/mol. Docking of the 22 non-covalent ligands to Mpro with AutoDock Vina was also performed (Fig.~\ref{fig:hist_vin_ad}). Vina scores are shifted to higher values and have a more one-sided and less disperse distribution than AutoDock-GPU scores. These differences reflect the different potentials and algorithms used in the two programs, and can be useful information when choosing a cut-off for score-based virtual screening.

 \subsubsection{Effect of box size and active site structure}
 
\begin{figure}[h]
     \centering
     \includegraphics[width=\linewidth]{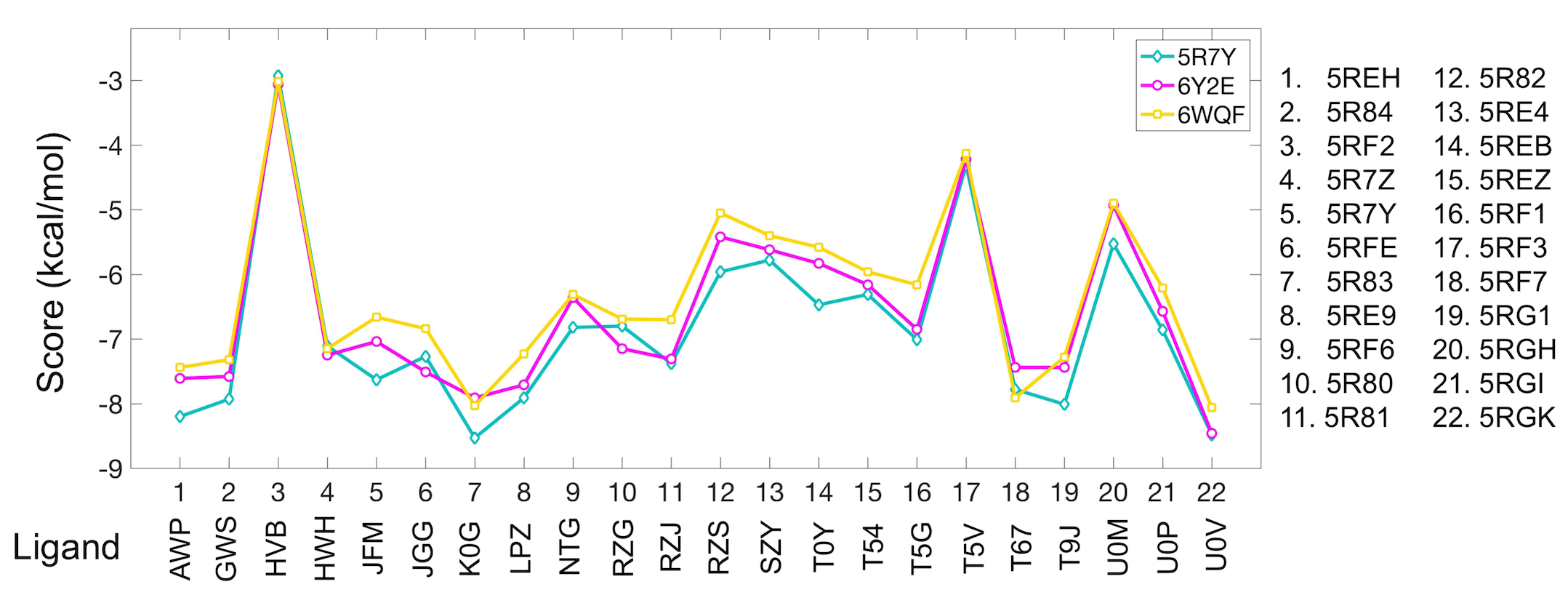}
     \caption{Scores obtained by binding the 22 crystallized non-covalent active-site bound ligands, using three different crystal structures for Mpro, each with a slightly different active site. Residue names given on x-axis and corresponding PDB IDs on left. All docking calculations were performed with SW using \texttt{nruns} = 100.}
    \label{fig:3scores}
 \end{figure}

 We docked the 22 active-site binders to several available Mpro crystal structures, PDB IDs: 5R7Y, 5R80, 5R81, 5R84, 5RE9, 5RF3, 5RGI, 6WQF, 6Y2E. Using a box that was fit very tightly about the active site, of dimensions 18.75$\times$24.75$\times$ 22.5 \AA, scores obtained were high-- not lower than -6.5 kcal/mol and with a mean of around -5. Using a box of size 26.25 \AA~ per side (with center as Pro 39 C-alpha atom) improved scores by approximately 2 kcal/mol, bringing scores down below -8 in several cases. We also noted that, using the same search box, the resulting scores were somewhat sensitive to small changes in the protein active site. These may include the position of several coils or small helix regions at the opening of the active site and flanking side chains. Fig.~\ref{fig:3scores} illustrates this difference, showing scores for docking the 22 active site binder ligands to two different Mpro structures, PDB IDs 6Y2E and 6WQF, which were crystallized without bound ligands, and 5R7Y, crystallized with a bound ligand.
 
\begin{figure}
    \centering
    \includegraphics[width=0.9\columnwidth]{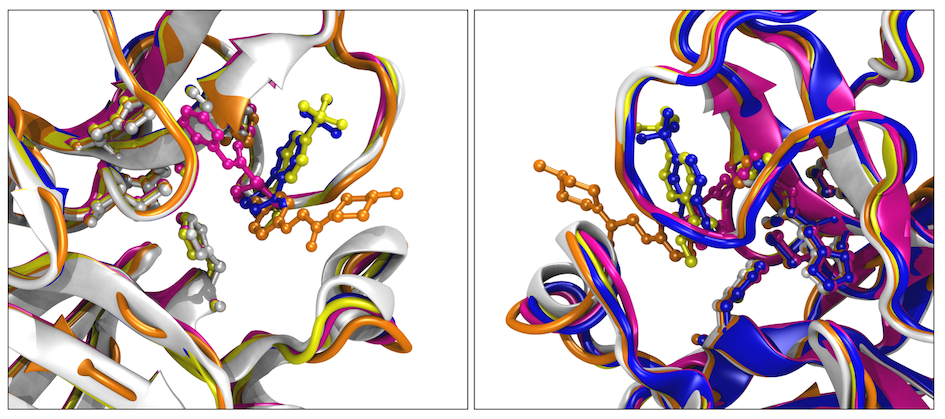}
    \caption{Overlay of the five experimental crystal structures of Mpro which are bound to non-covalent ligands showing the most variation in active site region conformation. Left and right panels show views rotated 180 degrees.}
    \label{fig:5most}
\end{figure}

 6Y2E and 6WQF differ by a large change in position of a side chain of the GLN 189 residue, seen upon close inspection of the 6WQF active site. This conformation, in addition to small changes in the positions of surrounding loops, gives the active site a reduced volume, thus limiting the search space for docking and potentially contributing to the slightly higher energy values obtained from docking to this structure. Fig.~\ref{fig:5most} shows an overlay of five of the 22 structures crystallized with ligand bound, that were found to be most different from the others (PDB IDs 5RGK, 5R84, 5R80, 5R81, 5RE9), and also from the unbound Mpro structures. Notable is the difference in the small helix that becomes a coil in several structures, and adopts a wide range of positions, changing the active site volume.
 
\subsection{Re-docking results: poses}
\begin{figure}[h]
    \centering
    \includegraphics[width=0.6\columnwidth]{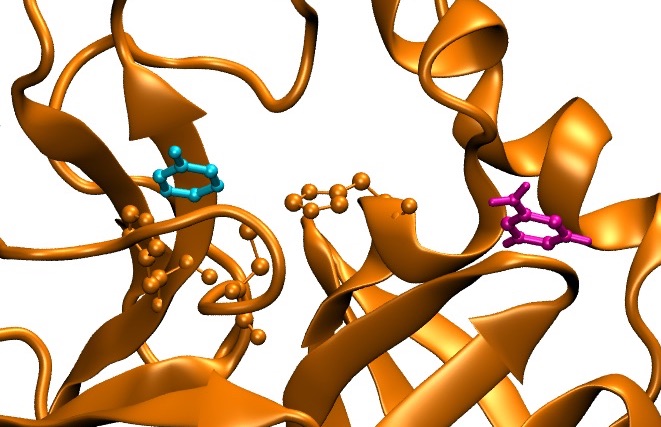}
    \caption{PDB 5RF3 shown in orange cartoon with crystallized T5V ligand, shown in cyan (CPK), along with three key residues (HIS 41, CYS 145, HIS 163) found in the active site shown in orange with atoms visible (CPK). Docked position shown in magenta in CPK representation. }
    \label{fig:T5V}
\end{figure}

\begin{figure}
    \centering
    \includegraphics[width=0.6\columnwidth]{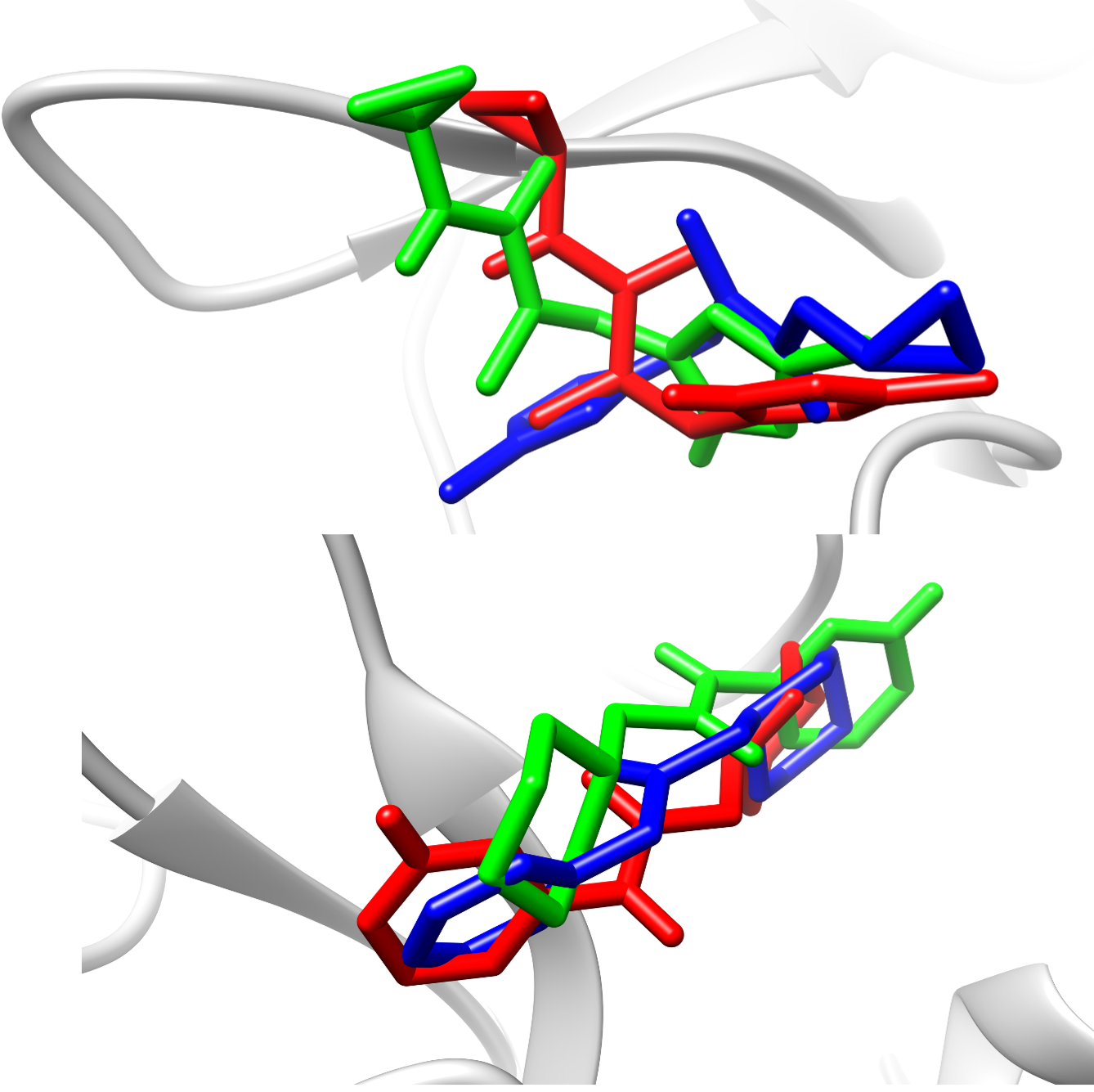}
    \caption{Two re-docking results comparing AutoDock-GPU and AutoDock Vina using PDB IDs 5rgi (top panel) and 5r84 (bottom panel). Vina: green; AutoDock-GPU: red; crystal structure: blue.}
    \label{fig:redock_comp}
\end{figure}

Re-docking refers to the docking of a ligand, crystallized in complex with a receptor, to the same receptor. We performed re-docking experiments on four of the 22 non-covalent binders, ligands from PDB IDs 5R7Y, 5R84, 5RF3, and 5RGI. After docking, the top-ranked (lowest score) docked pose of the ligand was superimposed on the crystal structure ligand pose to compare them. 5RF3 (ligand residue name T5V) is a very small fragment and can dock in many locations in the large active site of Mpro, this is shown in Fig.~\ref{fig:T5V}; it also receives a high value for the binding free energy, -4.75 kcal/mol. For two of the four ligands tested, PDB IDs 5R84 (ligand residue name GWS) and 5RGI (ligand residue name U0P), re-docking results were remarkably good, as illustrated in \ref{fig:redock_comp}. 5R7Y was docked in a less accurate position than these two.

\begin{figure}
    \centering
    \includegraphics[width=0.7\columnwidth]{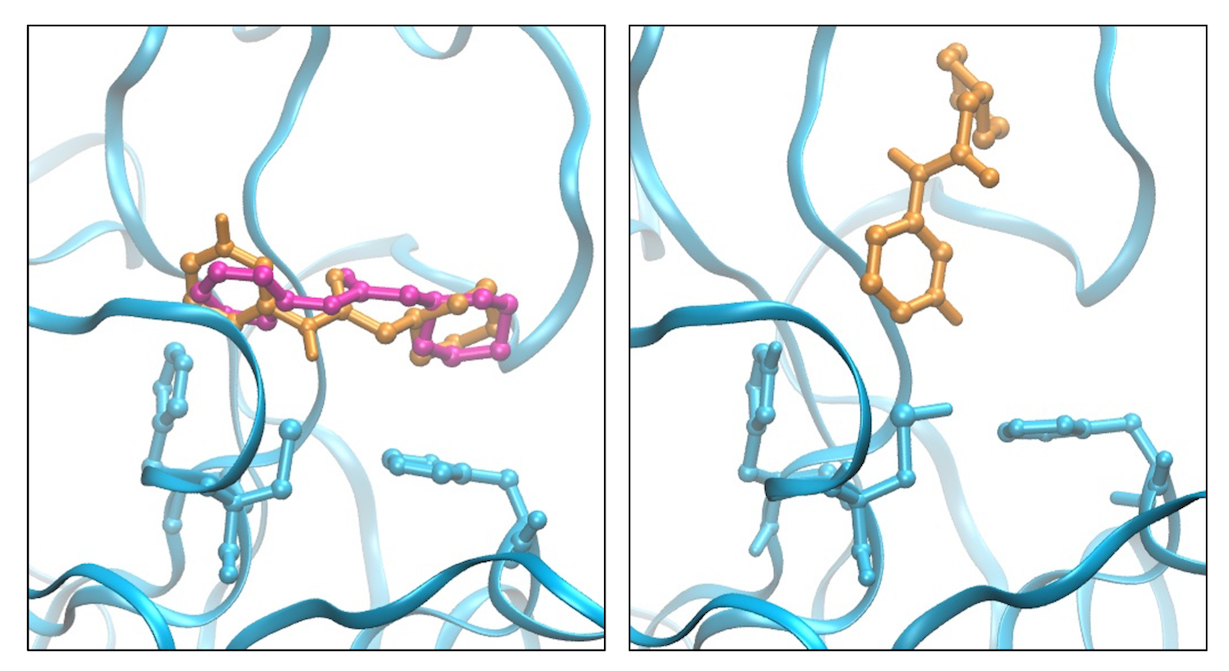}
    \caption{Two re-docking results (orange CPK) for the GWS ligand with Mpro stucture 5R84 (cyan ribbons), using a search box with volume 26.25 \AA~ per side (left panel), and 18.75$\times$24.75$\times$ 22.5 \AA~ per side (right panel). Crystallographic ligand position also shown in magenta in left panel. Three key residues (HIS 41, CYS 145, HIS 163) found in the active site shown in cyan with atoms visible (CPK) in both panels.}
    \label{fig:GWS_comp}
\end{figure}

Fig.~\ref{fig:redock_comp} also shows a comparison to the same re-dockings with Vina, showing that AutoDock-GPU can calculate redocked poses of comparable closeness to the crystallized position as Vina. Using a smaller search box could help to better locate very small ligands such as T5V, however, the smaller search box described above prevented the GWS ligand from being docked correctly, as shown in Fig.~\ref{fig:GWS_comp}. This demonstrates some of the difficulties encountered in high-throughput docking, as it is difficult to find a search box that can ensure the best docking result for all ligand sizes, even with a very reduced subset containing only 4 torsions or fewer.

\section{Conclusions and Future Work}
We have presented a new version of AutoDock-GPU that has been ported to CUDA and containing a number of optimizations to enable more efficient processing of thousands of ligands per receptor to facilitate high-throughput structure-based \textit{in silico} drug discovery. This version has enabled AutoDock-GPU to be deployed on the Summit supercomputer, which opens up this resource for massive computational efforts for therapeutics, especially for combating the current COVID-19 pandemic. The total speedup for flexible docking of small compounds with a moderate number of search iterations, using the new pipeline is close to 10$\times$ relative to the previous single receptor/ligand workflow. This pipeline will be back-ported to the OpenCL version in future work, allowing this method to also be employed on non-NVIDIA systems. For deployment on Summit, next steps will involve integrating this program into workflow management systems to efficiently manage large scale docking campaigns.

\begin{acks}
This research was sponsored by the Laboratory Directed Research and Development Program at Oak Ridge National Laboratory (ORNL), which is managed by UT-Battelle, LLC, for the U.S. Department of Energy (DOE) under Contract No. DE-AC05-00OR22725, and used resources of the Oak Ridge Leadership Computing Facility, which is a DOE Office of Science User Facility supported under Contract DE-AC05-00OR22725. The development of the AutoDock-GPU was supported by the
National Institutes of Health (GM069832). The authors thank Jonathan Lefman and Geetika Gupta (NVIDIA) for essential coordination and  communication support and collection of important feedback.
\end{acks}

\bibliographystyle{ACM-Reference-Format}
\bibliography{ad-gpu2_acm_bcb,ACM-BCB}

\end{document}